# EFFICIENT COMPRESSION OF PROLOG PROGRAMS

**Alin-Dumitru Suciu**
Technical University of Cluj-Napoca, Department of Computer Science
26-28, George Barițiu St., RO-3400, Cluj-Napoca, Romania
Phone/Fax: +40-64-194491, Email: Alin.Suciu@cs.utcluj.ro
**Kálmán Pusztai**
Technical University of Cluj-Napoca, Department of Computer Science
26-28, George Barițiu St., RO-3400, Cluj-Napoca, Romania
Phone/Fax: +40-64-194491, Email: Kalman.Pusztai@cs.utcluj.ro

**Abstract**

We propose a special-purpose class of compression algorithms for efficient compression of Prolog programs. It is a dictionary-based compression method, specially designed for the compression of Prolog code, and therefore we name it PCA (Prolog Compression Algorithm). According to the experimental results this method provides better compression than state-of-the-art general-purpose compression algorithms. Since the algorithm works with Prolog syntactic entities (e.g. atoms, terms, etc.) the implementation of a Prolog prototype is straightforward and very easy to use in any Prolog application that needs compression. Although the algorithm is designed for Prolog programs, the idea can be easily applied for the compression of programs written in other (logic) languages.

**1. Introduction**

The need for compression of Prolog programs naturally occurs in practice whenever we need to send Prolog code over the network (e.g. in distributed logic frameworks, mobile logic agents etc.) or to store large libraries of Prolog modules. Note that compression could be used also as a form of protection if the adversary does not know the compression algorithm.

Whenever the need for such compression appears one can choose to use a general-purpose compression algorithm [4] (e.g. Huffman, LZW, etc.) or to design and use a special-purpose compression algorithm that best suits for one's needs. Although designing a new compression algorithm requires some effort it is however expected from a special-purpose compression algorithm to achieve a better compression ratio that will justify the effort.

We will present a special-purpose class of compression algorithms for efficient compression of Prolog programs. It is a dictionary-based compression method, specially designed for the compression of Prolog code, and therefore we name it PCA (Prolog Compression Algorithm). According to the experimental results this method provides better compression and is faster than state-of-the-art general-purpose compression algorithms. Since the algorithm works with Prolog syntactic entities (e.g. atoms, terms, etc.) the implementation of a Prolog prototype is straightforward and very easy to use in any Prolog application that needs compression. Although the algorithm is designed for Prolog programs, the idea can be easily applied for the compression of programs written in other (logic) languages. There are two natural questions to answer when we talk about the compression of Prolog code:
1. Do we need the comments any further?
2. Do we need the variable names any further?

According to the logic programming theory [3] and the Prolog Standard [2] if we remove the comments and rename the variables this will lead to a perfectly equivalent, although less readable, Prolog program. This action is known to every Prolog programmer by means of the "listing" predicate. It is a common situation in practice, which can greatly benefit from the compression algorithm we propose. According to the answer to the above questions, the table below summarizes the PCA family of compression algorithms:

|  | Remove comments | Don't remove comments |
|---|---|---|
| Rename variables | $PCA_0$ | $PCA_1$ |
| Don't rename variables | $PCA_2$ | $PCA_3$ |

Table 1: The PCA family of compression algorithms

Although from the information theory viewpoint the above algorithms are lossy (comments, original formatting and original variable names may be lost), the Prolog program obtained after compression and decompression is equivalent to the original one. It is obvious that removing the comments will provide a better compression and it is expectable that renaming the usually long names of the variables with shorter ones (e.g. A-Z) will also provide a better compression. Therefore it is expectable that $PCA_0$ will provide the best compression while $PCA_3$ will provide the worst compression.

We will concentrate on the $PCA_0$ algorithm which is detailed in section 2 while in section 3 the experimental results are shown; finally we draw some conclusions and present some further work issues.

## 2. The $PCA_0$ Compression Algorithm

From a logical point of view and for an easier understanding we can split the algorithm in five sequential steps, STEP 0..4, as one can see in Figure 1. The original Prolog program (PP) is successively transformed in a series of normal forms, $NF_{0..4}$, ($NF_0$ is the equivalent of a "listing" command); each normal form has an associated dictionary, $D_{1..4}$, and header, $H_{1..4}$. The five steps of the algorithm are detailed below.

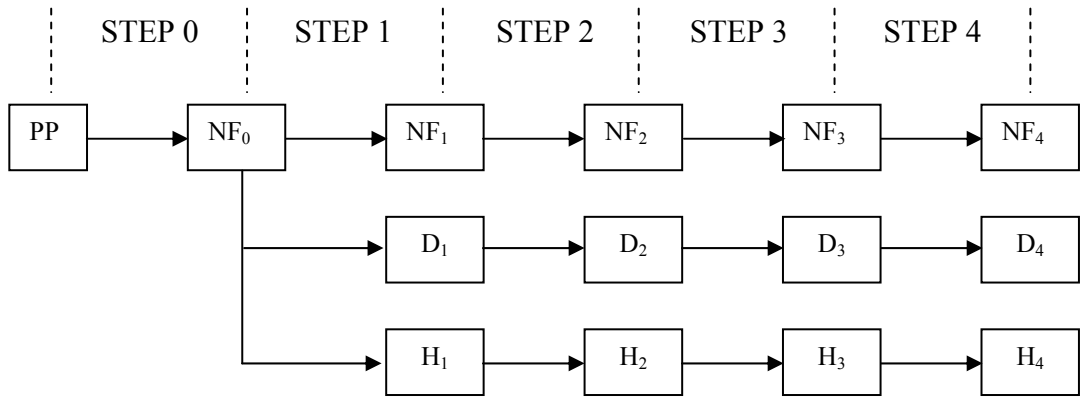

Figure 1: The $PCA_0$ Compression Algorithm

*STEP0 – comments removal and variable renaming*
In: Prolog Program (PP)
Out: Normal Form $NF_0$
Action:
This step is the equivalent of a "listing" command upon the original Prolog program. Comments and white spaces are removed, variables from each clause or directive are renamed from A to Z; in case we need more variables we use the naming convention A1, ..., A9, B1, ..., Z9, .... This is the only transformation of the algorithm that induces a loss of information and is therefore irreversible; at decompression the original program will not be recoverable, but the normal form $NF_0$ is perfectly equivalent from a computational viewpoint and can be used instead. In case someone needs the original variable names and/or the comments one may use the other compression algorithms from the PCA family.

*STEP 1 – building the dictionary and applying the substitution*
In: Normal Form $NF_0$
Out: Normal Form $NF_1$, Dictionary $D_1$, Header $H_1$
Action:
This is the essential transformation of the algorithm in which the first dictionary ($D_1$) is built; every lexical entity of the program (atom, functor, number, variable, constant, etc) has an entry in the dictionary; each entry consists of the name, arity and type (prefix, infix, postfix) of the lexical entity. The first header ($H_1$) is also built in this phase and contains information about the compression method that is needed at decompression (e.g. the size of the dictionary). The normal form $NF_1$ is obtained by replacing every lexical entity by its index in the dictionary and removing all the parentheses and commas. Thus the term p(a,B,f(c,d,e)) becomes &p&a&B&f&c&d&e where &x denotes the index of entity "x". Starting with this point the normal form becomes a binary representation of the program; indeed, suppose that for the example above the indexes are 102 56 42 79 100 101 27; if we maintain an ASCII representation then we get in fact an expansion (23 bytes from the original 15 bytes); on the contrary, if we switch to a binary representation, every index takes one byte so we get 7 bytes from the original

15 bytes. Depending on the size of the dictionary, the index will take one or more bytes; for a regular Prolog program two bytes should generally suffice.

*STEP 2 – more compact representation of indexes*
In: $NF_1, D_1, H_1$
Out: $NF_2, D_2, H_2$
Action:
The dictionary remains the same ($D_2 = D_1$). $NF_2$ is obtained in a similar way as $NF_1$ but the representation of the indexes is compacted; if the dictionary has N entries then we only need $\log_2 N$ bits to represent an index. The binary representations of the indexes are appended and grouped in bytes; the last byte is padded with zero. The number N is added to the header.

*STEP 3 – optimizing the dictionary and the header*
In: $NF_2, D_2, H_2$
Out: $NF_3, D_3, H_3$
Action:
The program representation remains the same ($NF_3 = NF_2$). The dictionary is split in three parts, one for each column:

| Part N (Names) | Part A (Arities) | Part T (Types) |
|---|---|---|

Part N contains the names of the lexical entities separated by a white space. We can replace some frequently used operators or built in predicates with ASCII codes 0-31 or 128-255. An additional optimization is obtained if all the variables are at the beginning of the dictionary; if NVAR is the number of the variables then the first NVAR entries in Part N can be removed and NVAR is added to the header $H_3$.

If AMAX is the maximal arity, then Part A can be compressed by representing the arities on $\log_2 AMAX$ bits and adding AMAX to the header $H_3$. Part T contains only three codes: 0=prefix, 1=infix, 2=postfix. Postfix operators are rare in practice so we can use a flag TF to compress Part T even more; if TF = 0 then there are no postfix operators and types will be represented on one bit; if TF=1 two bits will be used.

*STEP 4 – additional compression*
In: $NF_3, D_3, H_3$
Out: $NF_4, D_4, H_4$
Action:
In case one needs an even better compression, the output of STEP 3 ($H_3+D_3+NF_3$) can be further compressed using traditional compression algorithms or even external compression programs.

Finally we mention that an efficient implementation of the algorithm will combine and rearrange actions from different steps and will not generate all the intermediate normal forms but only the final result.

## 3. Experimental Results

We developed a prototype implementation of the algorithm $PCA_0$ in SICStus Prolog that allowed the study of the compression ratio for a set of test examples; compression time was not the focus of our research yet, as this will require a more efficient implementation. As test examples we used some of the source files from the SICStus Prolog Library.

Tables 2 and 3 below show the evolution of compression ratio during the steps of the algorithm (Ratio x corresponds to STEP x). In Table 2 the original size of the logic program PP is the reference while in Table 3 the normal form $NF_0$ (i.e. the original program with comments striped and the variables renamed) is the reference. As expected, the first table shows better results, the extra compression obtained is due to the comments removal.

| File | Ratio 0 | Ratio 1 | Ratio 2 | Ratio 3 | Ratio 4 |
|---|---|---|---|---|---|
| arrays.pl | 0.606 | 0.555 | 0.472 | 0.225 | 0.203 |
| assoc.pl | 0.453 | 0.422 | 0.432 | 0.184 | 0.135 |
| db.pl | 0.654 | 0.660 | 0.544 | 0.262 | 0.149 |
| atts.pl | 0.767 | 0.736 | 0.703 | 0.280 | 0.227 |
| heaps.pl | 0.463 | 0.448 | 0.391 | 0.188 | 0.147 |
| objects.pl | 0.465 | 0.598 | 0.407 | 0.262 | 0.209 |
| sockets.pl | 0.781 | 0.812 | 0.613 | 0.318 | 0.175 |

Table 2: Compression ratio with respect to the size of PP

| File | Ratio 1 | Ratio 2 | Ratio 3 | Ratio 4 |
|---|---|---|---|---|
| arrays.pl | 0.916 | 0.778 | 0.371 | 0.335 |
| assoc.pl | 0.931 | 0.954 | 0.406 | 0.298 |
| db.pl | 1.008 | 0.832 | 0.401 | 0.228 |
| atts.pl | 0.958 | 0.916 | 0.365 | 0.295 |
| heaps.pl | 0.968 | 0.843 | 0.405 | 0.318 |
| objects.pl | 1.285 | 0.874 | 0.564 | 0.450 |
| sockets.pl | 1.039 | 0.784 | 0.407 | 0.224 |

Table 3: Compression ratio with respect to the size of $NF_0$

| File | Winzip | Winrar | WinAce | $PCA_0$ |
|---|---|---|---|---|
| arrays.pl | 0.344 | 0.337 | 0.343 | 0.203 |
| assoc.pl | 0.282 | 0.278 | 0.280 | 0.135 |
| db.pl | 0.254 | 0.252 | 0.254 | 0.149 |
| atts.pl | 0.292 | 0.287 | 0.291 | 0.227 |
| heaps.pl | 0.332 | 0.329 | 0.331 | 0.147 |
| objects.pl | 0.375 | 0.369 | 0.377 | 0.209 |
| sockets.pl | 0.247 | 0.242 | 0.247 | 0.175 |

Table 4: Comparison with other programs (ratio with respect to the size of PP)

| File | Winzip | Winrar | WinAce | $PCA_0$ |
|---|---|---|---|---|
| arrays.pl | 0.568 | 0.555 | 0.567 | 0.335 |
| assoc.pl | 0.622 | 0.613 | 0.619 | 0.298 |
| db.pl | 0.388 | 0.386 | 0.388 | 0.228 |
| atts.pl | 0.381 | 0.374 | 0.380 | 0.295 |
| heaps.pl | 0.717 | 0.709 | 0.715 | 0.318 |
| objects.pl | 0.805 | 0.793 | 0.810 | 0.450 |
| sockets.pl | 0.316 | 0.310 | 0.316 | 0.224 |

Table 5: Comparison with other programs (ratio with respect to the size of $NF_0$)

Tables 4 and 5 above compare the compression ratio obtained by $PCA_0$ with other popular compression algorithms; we used Winzip 6.2, Winrar 2.04 and WinAce 0.96. Same as before, in Table 4 we considered the original program PP as the reference while in Table 5 the normal form $NF_0$ was taken as the reference. One can see that $PCA_0$ outperforms all the other compression programs for all the test examples. It is interesting to see that the difference between $PCA_0$ and the other programs is even bigger when there are no comments in the original program, which shows that the strength of the $PCA_0$ method does not lie in the comments removal but rather in the other original ideas used in the algorithm.

## 4. Conclusions and Further Work

We presented a class of special-purpose compression algorithms for Prolog programs named PCA (Prolog Compression Algorithm). The best algorithm of this class, $PCA_0$, provides better compression than state-of-the-art general-purpose compression algorithms.

Possible applications are in practice whenever we need to send Prolog code over the network (e.g. in distributed logic frameworks, mobile logic agents etc.) or to store large libraries of Prolog modules.

The prototype implementation we developed allowed us to study the performances of the algorithm $PCA_0$ only for the compression ratio; a more efficient implementation is under development that will allow the study of the compression and decompression time.

It is also a subject for further work the implementation of the rest of the algorithms from the PCA family and the development of similar special-purpose compression algorithms for other (logic) languages. Further refinements of the algorithms are also possible.

## References

[1] Apt, K.R.: From Logic Programming to Prolog, Prentice Hall, 1997.
[2] Deransart, P., Ed-Dbali, A., Cervoni, L.: PROLOG – The Standard, Springer, 1996.
[3] Lloyd, J.W.: Foundations of Logic Programming, Springer, 1984.
[4] Salomon, D.: Data Compression – The Complete Reference, Springer, 1998.
[5] Sterling, L., Shapiro, E.: The Art of Prolog, MIT Press, 1994.